\documentclass[a4paper,11pt]{article}
\pdfoutput=1 

\usepackage{jinstpub} 
\usepackage{comment}

\title{\boldmath AVOLAR. A high voltage generator for liquid argon time projection chambers}

\author[a,1]{L.~Romero,\note{Corresponding author.}}
\author[a]{J.M.~Cela,} 
\author[a]{E.~Sanchez Garcia,}
\author[a]{M.~Daniel,}
\author[b]{M.~de~Prado,}

\affiliation[a]{Centro de Investigaciones Energ\'{e}ticas, Medioambientales y Tecnol\'{o}gicas (CIEMAT), Av. Complutense 40, 28040 Madrid, Spain}
\affiliation[b]{Center for Proton Therapy, Paul Scherrer Institute, Villigen-PSI, Switzerland}

\emailAdd{luciano.romero@ciemat.es}

\abstract{Some of the main neutrino oscillation and dark matter experiments have chosen time projection chambers (TPC) filled with liquid argon (LAr) as their technology for the next generation of detectors. Because of its typical drift length of several meters, relatively large cathode voltages are desirable to provide a sizeable drift field. Current designs are based on feedthroughs with high voltages (HV) limited to several hundred kV. The present work proposes a novel method to produce higher voltages inside the detector.
It is based on a Van de Graaff HV generator where the charge transporting belt is replaced by a cryogenic LAr flow. Negative charge is injected in liquid by means of a grounded sharp point facing a positive voltage electrode with a high speed LAr stream in between. The LAr flow transports the charge to the cathode through an electrically insulating pipe. In the cathode the charge is extracted with a metallic mesh. The LAr flux is driven by a cryogenic helium pump with unidirectional valves assuring a continuous flow. The LAr operational temperature is maintained by a pressurized liquid nitrogen deposit with automatic filling. The whole system is installed within a dewar container that will be filled with LAr reproducing the typical TPC conditions.
This design has no mobile parts, so it is very robust and can be easily embedded within the structural support of a TPC cathode.
A prototype of this HV generator has been constructed at CIEMAT (Madrid), and is currently being characterized. This R\&D is presented and the preliminary results are discussed.}


\arxivnumber{2001.05268} 

\proceeding{LIDINE 2019: Light Detection in Noble Elements\\
28-30 August 2019\\
University of Manchester}

\begin{document}
\maketitle
\flushbottom

\section{Introduction}
\label{sec:intro}
The field of rare event physics requires of detectors with extremely high sensitivity and large volumes. Some of the main neutrino oscillation and dark matter experiments have chosen time projection chambers (TPC) filled with liquid argon (LAr) as their technology for the next generation of detectors. Example of those detectors are DUNE \cite{DUNE} and DarkSide-20k \cite{DarkSide}, both with drift lengths of several meters.

Ion/electron pairs are generated by particle interactions in  the active volume of a typical liquid argon TPC, the electrons and ions are pulled apart by means of a constant electric field $\vec{E_{d}}$. The drift field suppresses the recombination of the opposite charges created in the ionization track, reducing the so-called recombination light and making possible the collection of the ionization signal.

The electrons drifting toward the anode can be attached to electronegative impurities. Reducing the transit time of electrons, or, conversely, increasing the drift field, can reduce the requirements of LAr purity.

The background induced by noise and environmental radiation is taken during the data acquisition window, which, in turn, is defined by the maximum electron transit time. Therefore, an increment in the drift field reduce the background. 

The positive charges created in the LAr bulk have a mobility, thus a velocity, five orders of magnitude lower than the electrons \cite{Walkowiak:2000wf,Dey:1968,ICARUS:2015torti,atlas}. The electrons are promptly collected while the ions stay in the LAr for much longer time. As a consequence, in stable conditions, the average density of positive ions is much larger than that of electrons ($d_{I}  \gg d_{e}$). This space charge can locally modify the drift lines, the amplitude of the electric field, and ultimately the velocity of the  electrons, thus, a  displacement in the reconstructed position of the ionization signal can be produced. Additionally, recombination of drifting electrons with ions produced in previous interactions can cause a signal loss. 

Therefore, high electric drift fields can be beneficial for the operation of big LAr-TPC. However, to achieve a high drift field along a distance of several meters, very high voltages (HV) are necessary, which are currently limited to several hundred of KV by the electrical feedthroughs. The present work proposes a novel method to produce higher voltages inside the LAr-TPC detector.
AVOLAR is an acronym for "High voltage for LAr detectors". \footnote{In spanish:"Alto VOLtaje para detectores de ARgon l\'{i}quido".}

\section{AVOLAR system design.}
\label{sec:FE}

 The AVOLAR design is based on a Van de Graaff HV generator where the charge transporting belt is replaced by a cryogenic LAr flow. This concept was already patented in 1935 by Van de Graaff \cite{VandeGraaf:patent}. 
A liquid generator, using hexane, was implemented in 1969 \cite{Seeker:1969}, which reached 500 KV with an intensity of 30 $\mu$A.

 Negative charge is injected in a LAr by means of a grounded sharp point facing a positive HV flat electrode, with a high speed LAr flow in between, as shown in figure 1 left. The LAr is doped with oxygen, a electro-negative molecule which captures the electrons. The LAr flow transports the negative ions to the cathode through an insulating pipe. There the charge is extracted with a metallic mesh and the discharged LAr flow returns through a concentric external pipe, figure 1 right. The hydrodynamic LAr flow drags the charges if its velocity is larger than the velocity of the negative ions in the injection gap. This is easily achievable given the small value of the ion mobility in LAr\cite{Walkowiak:2000wf,Dey:1968,ICARUS:2015torti,atlas}. 

\begin{figure}
 \begin{center}
  \includegraphics[height=.35\textheight]{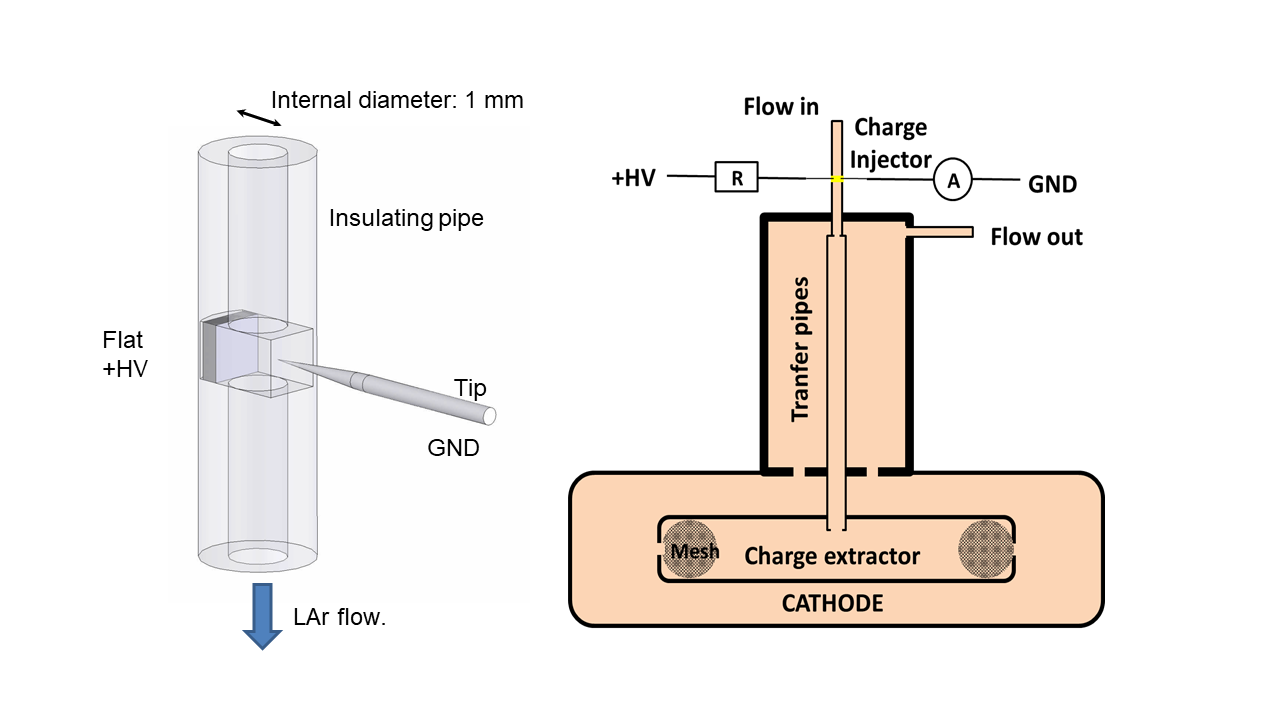}
  \caption{Left: Concept of charge injection. Right: Diagram of the LAr Van de Graaff generator.}
  \label{COM_UG}
  \end{center}
\end{figure}
\begin{figure}
 \begin{center}
  \includegraphics[height=.35\textheight]{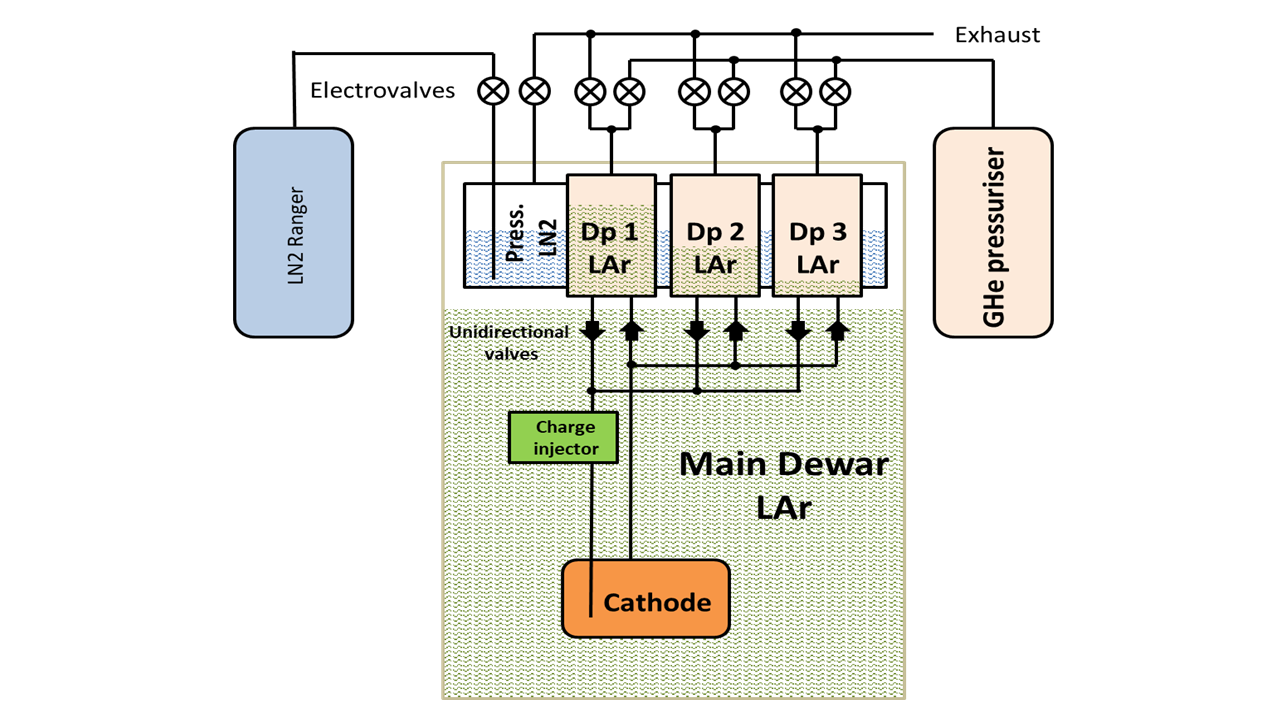}
  \caption{Schematic of AVOLAR global system, not to scale.}
  \label{COM_UG1}
  \end{center}
\end{figure}

The injector and cathode assembly is submerged in  LAr, which acts as an insulator, and hangs from a drum filled with liquid nitrogen (LN$_{2}$) which is the cooling agent, see the figure 2. The LN$_{2}$ is pressurized within the drum to raise its equilibrium temperature to the level of the LAr at 1 bar. Embedded in this drum there are 3 cylindrical deposits that hold the recirculating LAr and are alternatively used as a source and destination of LAr. High pressure helium gas pushes the LAr through the injector and the cathode. Unidirectional valves make the flow continuous and with the correct direction. 

\section{AVOLAR implementation and preliminary tests}
\label{sec:FL}

A prototype of AVOLAR was build in the mechanical workshop of CIEMAT. It is shown in figure 3. This prototype is introduced in a main dewar of 60 cm in diameter and 120 cm in height.

The gas system of the prototype has five instrumented volumes, the main dewar volume, the LN$_{2}$ drum and the three deposits for the recirculating LAr. Each volume has an electronic pressure gauge, a capacitive level transducer and a temperature sensor. The system has lines for the supply of helium, argon and vacuum. Safety of the gas system is enforced by spring actuated venting valves in each of the volumes. The cryogenic system is based on a 180 liters external LN$_{2}$ ranger. The LN$_{2}$ drum is automatically filled and its pressure is kept between 2 and 2.6 bar by means of two electrovalves driven by an Arduino control card. Filling of the recirculating LAr is done condensing argon gas in the walls of the 3 deposits within the LN$_{2}$ drum. Before condensing, a pressure of 100 mbar of oxygen is established in the LAr deposits to dope the LAr. The time needed for cooling and filling AVOLAR is around 4 hours.

\begin{figure}[t!]
\begin{center}
  \includegraphics[height=.35\textheight]{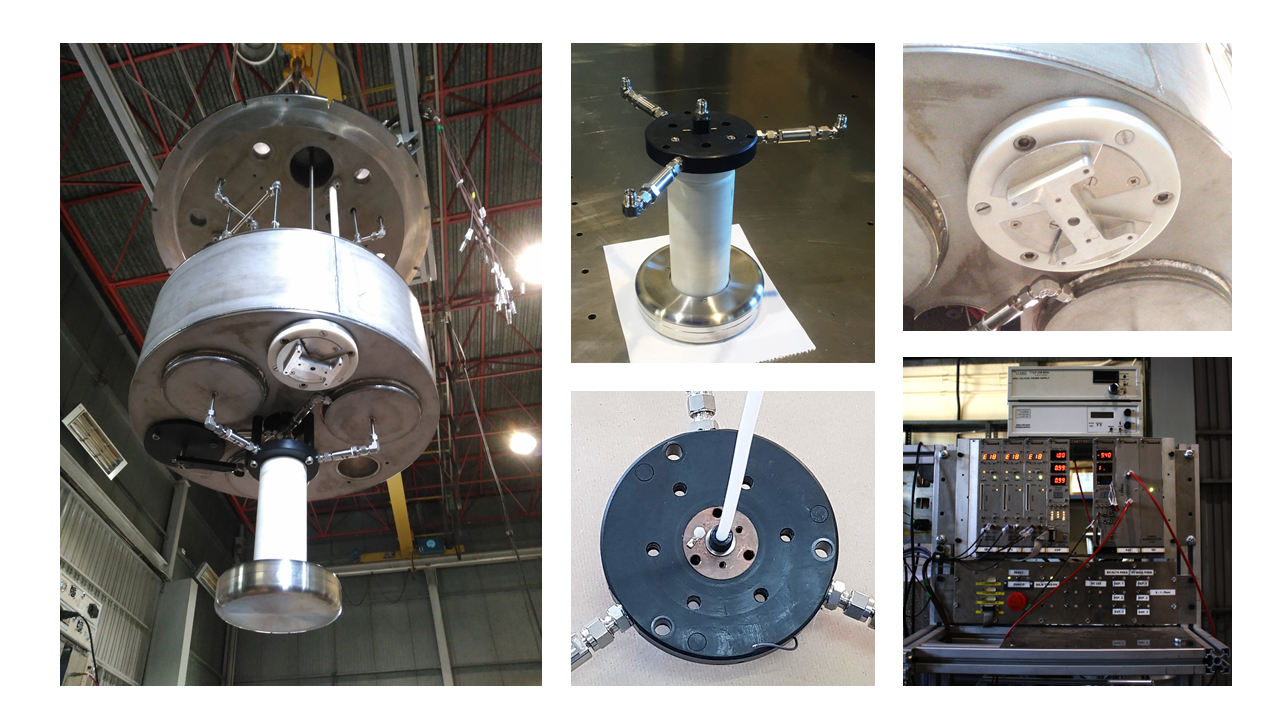}
  \caption{Left: AVOLAR prototype out of the main dewar. Center up: The injector and cathode assembly. Center down: The injector with the guard electrode and the central pipe. Right up: Generating voltmeter. Right down: Control electronics. }
  \label{Field_lines}
  \end{center}
\end{figure}

To operate AVOLAR there are 2 valves per recirculating deposit, one to apply pressurized helium and a second to connect the deposit to the vent line. Currently those valves are manually operated, however, we plan to automatize them. The helium gas used to push the LAr through the injector has a pressure of 8 bars. With this pressure the LAr speed in the 1 mm diameter injector passage was obtained measuring the LAr volume transferred in a given time. It was estimated to be 30 m/s.

To do the electrical test, the sharp point of the injector was connected to ground through an ammeter. The flat electrode is set to a positive high voltage, up to 15 kV, through a 20 Gigaohm resistor. A retractable probe allows electrical access to the cathode. Through this probe, the cathode is connected to ground through an ammeter. A guard ring around the external side of the injector prevents the build up of stray charge. This electrode is shown in figure 3, center down.

Initial tests demonstrated charge injection in the LAr flow. In these first tests, the charge leaked out of the internal pipe, being recovered in the guard ring. Figure 4-left shows the positive current out of the tip and the negative current going into the guard ring. The excitation voltage is 3 kV. Both currents are smooth and equal, they have a value of $\approx$ 200 nA. Small sparks to the guard ring can be observed and also an injection stop due to unknown reasons.

Once the electric charge leak was fixed, we transferred charge to the cathode. Figure 4-right show the positive current going out of the point and the negative current going into the cathode. Both currents are equal. The voltage across the electrodes was 8 kV and the steady electric current was 200 nA. Two fast positive pulses in both currents are attributed to discharges. Those discharges erode the point, making it necessary to, gradually, rise the voltage between electrodes. We are working in geometries more robust to discharges, in particular a razor blade instead of a point.

The HV value of the cathode is measured by a generating voltmeter. This instrument is shown in figure 3, top right. An electrode, exposed to the cathode electric field, is covered and uncovered alternatively by a grounded plate, generating a periodic signal of induced charge. The amplitude of the signal is proportional to the cathode voltage. The generating voltmeter was calibrated setting a known voltage in the cathode via the retractable probe.

%
High voltage test were done retracting the probe from the cathode. Currently the main Dewar is filled with air so the HV achieved before sparking between the cathode and ground is limited. We measured 50 kV in the cathode using the generating voltmeter. We will fill the Dewar with LAr once we fully understand and optimize the charge transfer mechanism.

\begin{figure}[t!]
\begin{center}
  \includegraphics[height=.34\textheight]{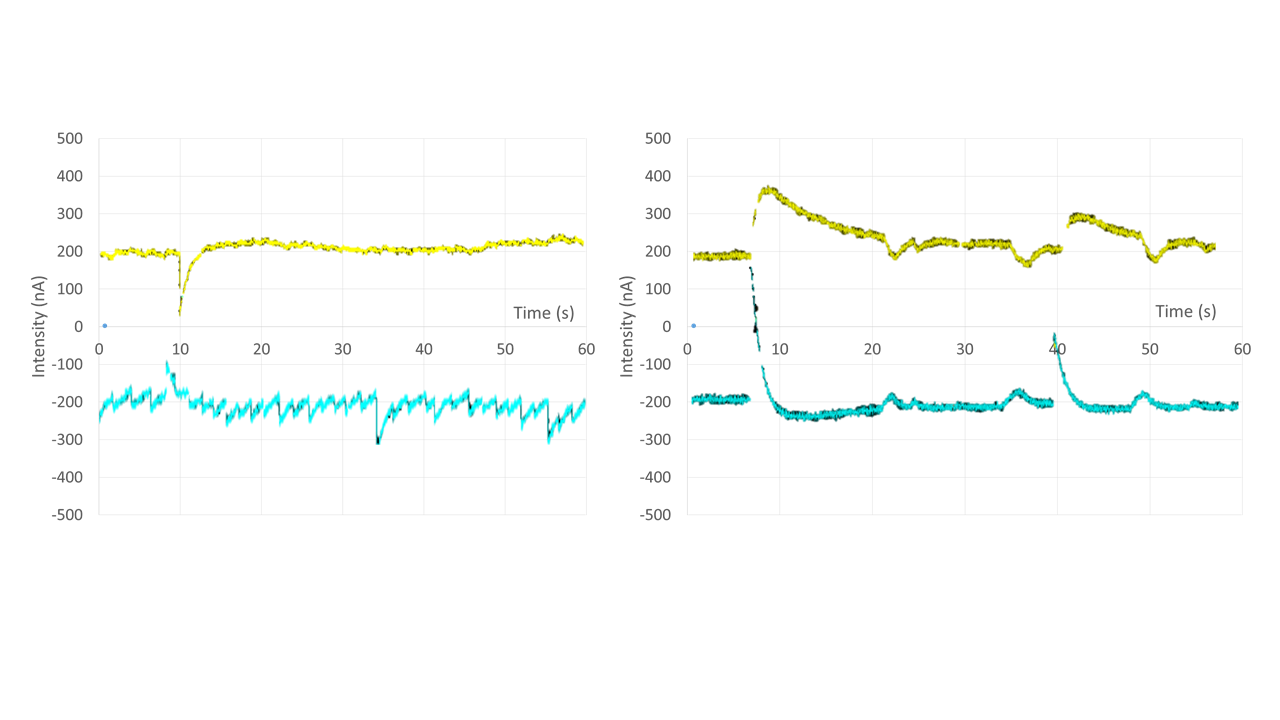}
  \caption{Left: Leak current in nA at the point and at the guard ring. The point HV is 3 kV. Right: Currents in nA at the point and at the cathode. The point HV is 8 kV. Time in seconds. }
  \label{Field_lines1}
  \end{center}
\end{figure}

\section{Conclusions}
A research line is ongoing at CIEMAT to develop a high voltage generator based on a liquid argon Van de Graaff design. This generator would be of utility in massive liquid argon TPC detectors.

The fact that this HV generator operates at LAr temperature eases its integration in a big detector. Because the HV is internal, no bushing is needed, as opposed to the case of externally generated HV. The transfer tubes may be of any length, so they can be integrated in the mechanical support of the cathode. They can be fabricated in plastic with a high electrical rigidity, so the HV generator can be very robust against electrical discharges.

The HV generated would help to reduce the positive space charge and to minimize the background.

Transfer of charge from the injector to the cathode has been demonstrated. New approaches are being currently investigated to optimize the quality of the charge transfer and eliminate spurious discharges. The next step is filling the main dewar with LAr in order to raise the voltage in the cathode at least to 1 MV. The HV generated would be a step forward to bias the cathode and induce a high drift field in a massive detector.


\acknowledgments
This research is funded by the Spanish Ministry of Economy and Competitiveness (MINECO) through the grant FPA2017-82647P. The authors are also supported by the ``Unidad de Excelencia Mar\'{i}a de Maeztu: CIEMAT - F\'{i}sica de Part\'{i}culas'' through the grant MDM-2015-0509.


\end{document}